\documentclass[final,3p,times]{elsarticle}

\usepackage{amssymb}
\usepackage{amsmath}

\begin{document}

\begin{frontmatter}


\title{Semileptonic decay form factors of $\Xi_b^0 \rightarrow \Xi_c^+\ell\bar{\nu}_{\ell}$ in HQET}


\author[1,2]{Kinjal Patel}
\ead{patelkinjal.dpphd23@vnsgu.ac.in}

\author[1,2]{Kaushal Thakkar}
\ead{kaushal2physics@gmail.com}

\affiliation[1]{organization={Department of Physics, Government College Daman},
                city={Daman},
                postcode={396210},
                state={U. T. of Dadra $\&$ Nagar Haveli and Daman $\&$ Diu},
                country={India}}


\affiliation[2]{organization={Veer Narmad South Gujarat University},
                city={Surat},
                state={Gujarat},
                country={India}}

\begin{abstract}
Heavy-to-heavy semileptonic decays, particularly the bottom-to-charm quark transitions, are essential for testing the Standard Model (SM) and extracting the Cabibbo-Kobayashi-Maskawa (CKM) matrix elements. These decays have been extensively studied using various theoretical approaches. In this work, we investigate the semileptonic decay $\Xi_b^0 \rightarrow \Xi_c^+\ell\bar{\nu}_{\ell}$ (where $\ell = e$, $\tau$) using a phenomenological quark model. We compute the ground-state masses of the initial and final baryons to get the wave function, which is then used to calculate the form factors, including corrections up to order $1/m_Q$ within the framework of Heavy Quark Effective Theory (HQET). The obtained form factors are implemented in the helicity formalism to evaluate the differential decay rates, total decay width and branching ratio. We compare our results for the form factors at both the maximum and minimum recoil points with previous theoretical studies, finding good agreement. We observe that the form factors depend on the transferred momentum $q^2$ and their magnitude gradually increases with increasing $q^2$. The dominant form factors are $f_1$ and $g_1$, and they also exhibit similar $q^2$ dependencies. Additionally, we calculate the lepton flavour universality (LFU) ratio $R(\Xi_c) \approx 0.3$, which is in agreement with existing theoretical predictions.
\end{abstract}


\end{frontmatter}


\section{Introduction}
In the last few years, significant experimental progress has been made in the study of the properties of heavy baryons. Heavy baryons containing a bottom quark serve as an excellent platform for exploring the interplay of weak and strong interactions within the Standard Model. Their internal structure, involving one heavy bottom quark and two lighter quarks, allows for a clean separation between short-distance weak processes and long-distance QCD effects. This makes them ideal systems for probing heavy quark symmetry, studying non-perturbative QCD dynamics, and understanding the mechanisms behind semileptonic decays.
As of now, approximately 32 charmed and 30 bottom-flavoured singly heavy baryons are listed in the Particle Data Group (PDG) \cite{pdg2024}. The $\Lambda_c^+$ was first reported by Fermilab in 1976 \cite{Knapp1976}. The production of $\Lambda_b^0$ baryon was confirmed early at CERN Intersecting Storage Rings (ISR) and later reported by several collaborations \cite{Bari1991a,Bari1991b}. The two states $\Sigma_c^{+*}$ and $\Sigma_c^+$ were observed in their $\Lambda^+_{c}\pi^0$ decay \cite{Ammar2001}. The strange-bottom baryon $\Xi_b^-$ was observed by D$\varnothing$ experiment by proton-antiproton collision at Fermilab \cite{Abazov2007}. The excited bottom-strange states $\Xi_b(6327)^0$, $\Xi_b(6333)^0$, $\Xi_b(6100)^-$ and $\Xi_b(6227)^0$ have been reported by LHCb and CMS collaboration \cite{Aaij1,Aaij2,Sirunya2021}. $\Sigma_b^0$, $\Sigma_b^{0*}$ and $\Omega_b^{-*}$ baryons have not yet been confirmed. So far,the $\Lambda_b^0$ baryon has been more widely studied than other bottom baryons, including $\Xi_b^0$ and $\Xi_b^-$. Two decay modes, $\Xi_b^0 \rightarrow \Xi_c^+ D_s^-$ and $\Xi_b^- \rightarrow \Xi_c^0 D_s^-$ have been observed for the first time using proton-proton collision data collected by the LHCb experiment \cite{RAaij2023epjc}, which indicates that the $\Xi_b^- \rightarrow \Xi_c^0 \ell\bar{\nu}_{\ell}$ transition may soon be conducive to experimental observation.\\
Theoretically, the $\Xi_b^- \rightarrow \Xi_c^0 $ transition has been studied using various theoretical methods, including nonrelativistic Quark model (NRQM) \cite{Cheng1996,Hassanabadi2015,Albertus2005,Patel2025a}, Lattice QCD \cite{Bowler1998}, Relativistic Quark Model (RQM) \cite{Ivanov2000,Ebert2006,Migura2006,Ivanov1997,Faustov2018}, QCD sum rules (QCDSR) \cite{Zhao2020,Neishabouri2025}, Bethe-Salpeter Equation \cite{Ivanov1999}, and, light front approaches \cite{Ke2024,Zhao2018,Cardarelli1999}. In this study, we extend our previous work on the semileptonic transition of $\Lambda_b^0$ baryon \cite{Thakkar2020}. The present work provides a consistent analysis of the semileptonic decay $\Xi_b^0 \rightarrow \Xi_c^+\ell\bar{\nu}_{\ell}$ within the framework of the Hypercentral Constituent Quark Model combined with Heavy Quark Effective Theory (HQET). In this approach, the baryon wavefunctions are utilised to evaluate the Isgur-Wise function and its associated parameters, which are subsequently employed to compute all six form factors, including subleading $1/m_Q$ corrections. The Lepton Flavour Universality ratio $R(\Xi_c)$ is also computed in the present study, which allows for a direct comparison between different lepton channels and highlights the role of lepton-mass-dependent effects in the decay dynamics.

\section{Theoretical Framework}\label{sec:1}
The decay properties of $\Xi_b^0$ baryon are studied within the framework of Hypercentral constituent quark model (HCQM). This well-established model is effective for describing the internal dynamics and properties of baryons \cite{Ferraris1995,Giannini1983}. In HCQM, Jacobi coordinates are essential for simplifying the three-body problem, providing a simplified representation of inter-quark dynamics. The internal dynamics are described using Jacobi coordinates with the collective hyper radius (which contains three-body effects) defined as $x=\sqrt{\rho^{2}+\lambda^{2}}$. The six-dimensional hyperradial Schr\"{o}dinger equation can be written as
\begin{equation}\label{eq:1}
\left[-\frac{1}{2m}\frac{d^2}{dx^2} + \frac{\frac{15}{4}+\gamma(\gamma + 4)}{2mx^2} + V(x) \right] \phi_{\gamma}(x) = E\phi_{\gamma}(x),
\end{equation}
where $\phi_{\nu\gamma}=x^{\frac{5}{2}}\psi_{\gamma}(x)$ is the hyper-radial wave function, where $\psi_{\gamma}(x)$ is the hypercentral wave function labelled by the grand angular quantum number $\gamma$ defined by the number of nodes $\nu$. The potential is assumed to depend only on the hyper radius and hence is a three-body potential since the hyper radius depends only on the coordinates of all three quarks. The hyperCoulomb (hC) plus linear potential, which is given as
\begin{equation}\label{eq:2}
V(x) = \frac{\tau}{x} + \beta x  + V_0,
\end{equation}
where $\tau$ = $-$$\frac{2}{3}\alpha_s$ is the hyperCoulomb strength and the values of the potential parameters $\beta$ and $V_0$ are fixed to obtain the ground-state masses. $V_{spin}$ is the spin dependent part, which is perturbatively added, which is given as in \cite{Garcilazo2007,Patel2025}
\begin{equation}\label{eq:3}
V_{spin}(x) = -\frac{A}{4} \alpha_s  \frac{e^{-x/x_0}}{x {x_0}^2} \sum_{i<j} \boldsymbol{\lambda}_i \cdot \boldsymbol{\lambda}_j \frac{\boldsymbol{\sigma}_i \cdot \boldsymbol{\sigma}_j}{6 m_i m_j},
\end{equation}
here, the parameter $A$ and the regularisation parameter $x_0$ are considered as the hyperfine parameters of the model. The hyperfine parameters $A$ and $x_0$ are discussed in detail, see Ref. \cite{majethiya2008a}. The values of parameters are listed in Table \ref{tab:table1}. $\boldsymbol{\lambda_{i,j}}$ are the SU(3) colour matrices, $\boldsymbol{\sigma_{i,j}}$ are the spin Pauli matrices, and $m_{i,j}$ are the constituent masses of two interacting quarks. The masses of ground-state $\Xi_Q$ baryons are calculated by summing the model quark masses (see Table \ref{tab:table1}), kinetic energy, and potential energy.
\begin{equation}\label{eq:4}
M_B = m_1 + m_2 + m_3 + \langle H \rangle.
\end{equation}

\section{Form factors and semileptonic decay of $\Xi_b^0$ baryon}\label{sec:3}
The hadronic matrix elements of the vector and axial-vector currents can be parameterised in terms of six form factors as follows \cite{Gutsche2015,Patel2026}
\begin{align}\label{eq:5}
M_{\mu}^{V} = \langle\Xi_{c}^+|\bar{c}\gamma_{\mu}b|\Xi_{b}^{0}\rangle = \bar{u}_{\Xi_{c}^+}(p^{\prime},\lambda^{\prime})\left[\gamma_{\mu}f_{1}(q^{2})-i\sigma_{\mu\nu}q^{\nu}f_{2}(q^{2})+q_{\mu}f_{3}(q^{2}) \right] u_{\Xi_{b}^0}(p,\lambda), \\ \nonumber
M_{\mu}^{A} = \langle\Xi_{c}^+|\bar{c}\gamma_{\mu}\gamma_{5}b|\Xi_{b}^{0}\rangle = \bar{u}_{\Xi_{c}^+}(p^{\prime},\lambda^{\prime})\left[ \gamma_{\mu}g_{1}(q^{2})-i\sigma_{\mu\nu}q^{\nu}g_{2}(q^{2})+q_{\mu}g_{3}(q^{2}) \right]\gamma_{5}u_{\Xi_{b}^0}(p,\lambda),
\end{align}
where $\sigma_{\mu\nu}=\frac{i}{2}[\gamma_\mu,\gamma_\nu]$. $\bar{u}_{\Xi_c}(p',\lambda')$ and $u_{\Xi_b}(p,\lambda)$ are the Dirac spinors of the $\Xi_c$ and $\Xi_b$ baryons, and $p^{(')}$ and $\lambda^{(')}$ are the corresponding momentum and helicity, respectively. Another parameterisation of these decay matrix elements can be expressed as follows
\begin{align}\label{eq:6}
M^{V}_{\mu} = \bar{u}_{\Xi_c^+}(p',\lambda') \left[\gamma_{\mu}\,F_{1}(q^{2}) + F_{2}(q^{2})\,\frac{p_{\mu}}{M_{\Xi_{b}^0}}
 + F_{3}(q^{2})\,\frac{p'_{\mu}}{M_{\Xi_{c}^+}}\right]  u_{\Xi_b^0}(p,\lambda) \\ \nonumber
M^{A}_{\mu}, = \bar{u}_{\Xi_c^+}(p',\lambda')\left[ \gamma_{\mu}\,G_{1}(q^{2}) + G_{2}(q^{2})\,\frac{p_{\mu}}{M_{\Xi_{b}^0}}
+ G_{3}(q^{2})\,\frac{p'_{\mu}}{M_{\Xi_{c}^+}} \right] \gamma_{5}\,u_{\Xi_b^0}(p,\lambda),
\end{align}
$f_i$ and $g_i$ with $i=1,2,3$ are the three form factors that describe the vector and axial vector transitions, respectively. The relationship between these two sets of form factors, shown in Eq. (\ref{eq:5}) and (\ref{eq:6}), is as follows \cite{Falk1993}
\begin{align}\label{eq:7}
f_1 &= F_1 + (m_{\Xi_b} + m_{\Xi_c}) \left( \frac{F_2}{2m_{\Xi_b}} + \frac{F_3}{2m_{\Xi_c}} \right) \nonumber \\
f_2 &= -\frac{F_2}{2m_{\Xi_b}} - \frac{F_3}{2m_{\Xi_c}} \nonumber \\
f_3 &= \frac{F_2}{2m_{\Xi_b}} - \frac{F_3}{2m_{\Xi_c}} \nonumber \\
g_1 &= G_1 - (m_{\Xi_b} - m_{\Xi_c}) \left( \frac{G_2}{2m_{\Xi_b}} + \frac{G_3}{2m_{\Xi_c}} \right) \nonumber \\
g_2 &= -\frac{G_2}{2m_{\Xi_b}} - \frac{G_3}{2m_{\Xi_c}} \nonumber \\
g_3 &= \frac{G_2}{2m_{\Xi_b}} - \frac{G_3}{2m_{\Xi_c}}
\end{align}
The above relation between these two sets allows us to connect the HQET-based formulation with the helicity formalism used for numerical evaluation. In the framework of HQET, the structure of the form factors simplifies considerably in the infinite heavy-quark mass limit $m_Q\rightarrow\infty$ (Q = b,c). In this limit, heavy-quark spin symmetry implies that all form factors are reduced to a single universal Isgur-Wise function $\xi(\omega)$ \cite{Isgur1991}.

\begin{eqnarray}\label{eq:8}
F_1(q^2) = G_1(q^2) = \xi(\omega),\nonumber \\
F_2 = F_3 = G_2 = G_3 = 0,
\end{eqnarray}

where $\omega=\upsilon\cdot\upsilon'$ is the velocity transfer between the initial $\upsilon$ and the final $\upsilon'$ heavy baryons. This is related to the squared four-momentum transfer between the heavy baryons, $q^2$ as $\omega=\frac{m^2_{\Xi_b}+m^2_{\Xi_c}-q^2}{2m_{\Xi_b}m_{\Xi_c}}$. The Isgur-Wise function $\xi(\omega)$ is expanded as \cite{Thakkar2020}

\begin{equation}\label{eq:22}
\xi(\omega)=1-\rho^2 (\omega-1)+c(\omega-1)^2+\ldots
\end{equation}
This Taylor expansion of the Isgur-Wise function is carried out around the zero-recoil point ($\xi(\omega)\mid_{\omega=1} = 1$). Since the kinematic range remains close to this region in the present calculation, the expansion up to the quadratic order provides a valid approximation. $\rho^2$ is the magnitude of the slope and $c$ is the curvature (convexity parameter) of the IWF ($\xi(\omega)$), which can be written in HCQM as in \cite{Patel2025a,Thakkar2020}
\begin{equation}\label{eq:23}
\rho^2=16 \pi^2 m^2\int_{0}^{\infty} |\psi_{\nu\gamma}(x)|^2 x^7 dx,
\end{equation}
\begin{equation}\label{eq:24}
c=\frac{8}{3} \pi^2 m^4\int_{0}^{\infty} |\psi_{\nu\gamma}(x)|^2 x^9 dx.
\end{equation}
Beyond the heavy-quark limit, subleading corrections of the order $1/m_Q$ arise in HQET \cite{Falk1993,Neubert1994}. These corrections originate from two sources. The first one parameterises the $1/m_Q$ corrections to the HQET and is proportional to the product of the parameter $\bar{\Lambda} = m_{\Xi_b}-m_{b}$, which is the difference between the baryon mass ($\Xi_b$) and heavy quark mass ($b$) in the infinitely heavy quark limit and the leading order Isgur-Wise function $\xi(\omega)$. The second originates from the kinetic energy term in the $1/m_Q$ correction to the HQET Lagrangian and introduces the additional function $A(\omega) = \frac{\bar{\Lambda}}{1+\omega}\xi(\omega)$ \cite{Georgi1990}. The baryon form factors in the HQET are expressed as \cite{Falk1993,Georgi1990,Neubert1994}
\begin{eqnarray}\label{eq:19}
F_1(\omega) = \xi(\omega) + \left( \frac{1}{2m_b} + \frac{1}{2m_c} \right) \left[ B_1(\omega) - B_2(\omega) \right], \nonumber \\
G_1(\omega) = \xi(\omega) + \left( \frac{1}{2m_b} + \frac{1}{2m_c} \right) B_1(\omega), \nonumber\\
F_2(\omega) = G_2(\omega) = \frac{1}{2m_c} B_2(\omega), \nonumber\\
F_3(\omega) = -G_3(\omega) = \frac{1}{2m_b} B_2(\omega),
\end{eqnarray}
where the functions $B_1$ and $B_2$ are expressed as \cite{Falk1993}
\begin{eqnarray}\label{eq:20}
B_1(\omega) = \bar{\Lambda} \frac{w - 1}{w + 1} \, \xi(\omega) + A(\omega), \nonumber\\
B_2(\omega) = -\frac{2\bar{\Lambda}}{w + 1} \, \xi(\omega).
\end{eqnarray}

The helicity amplitudes for the $V-A$ currents as defined in the helicity formalism, expressed in terms of the baryon form factors \cite{Bialas1993}.
The helicity structure functions with definite parity can be expressed in terms of bilinear combinations of the helicity amplitudes, see Ref. \cite{Gutsche2015,Faustov2016,Patel2026} for details.
The differential decay rate is expressed as follows \cite{Gutsche2015,Faustov2016,Migura2006}
\begin{equation}\label{eq:11}
\frac{d\Gamma}{dq^2} = \frac{G_F^2}{8\pi^3}|V_{cb}|^2 \frac{\lambda^{\frac{1}{2}}(q^2-m_l^2)^2}{48M_{\Xi_b}^3 q^2} \mathcal{H}_{Total}
\end{equation}
where $G_F = 1.16\times10^{-5} GeV^{-2}$ is the Fermi coupling constant, $|V_{cb}|= 0.041$ is the CKM matrix element, the quantity $\lambda$ is defined as the standard $K\ddot{a}ll\acute{e}n$ function $\lambda = M_{\Xi_b}^4 + M_{\Xi_c}^4 + q^4 - 2(M_{\Xi_b}^2 M_{\Xi_c}^2 + M_{\Xi_c}^2 q^2 + M_{\Xi_b}^2 q^2) $, $m_l$ is the lepton mass ($l=e,\tau$). The total helicity is defined as follows
\begin{equation}\label{eq:12}
\mathcal{H}_{Total}(q^2) = \mathcal{H}_T(q^2) + \mathcal{H}_L(q^2) +\frac{m_l^2}{2q^2} (\mathcal{H}_T(q^2) + \mathcal{H}_L(q^2) + 3\mathcal{H}_S(q^2)),
\end{equation}
where the first two terms are non-spin-flip contributions and the last three terms proportional to $m_l$ are lepton-helicity flip contributions. By integrating the differential decay rate in Eq. (\ref{eq:11}) ($q^2 \in [m^2_e,(M_{\Xi_b}-M_{\Xi_c})^2]$), we obtain the total decay rate.

\section{Results and discussions}\label{sec:3}
\begin{figure}
\includegraphics[scale=0.3]{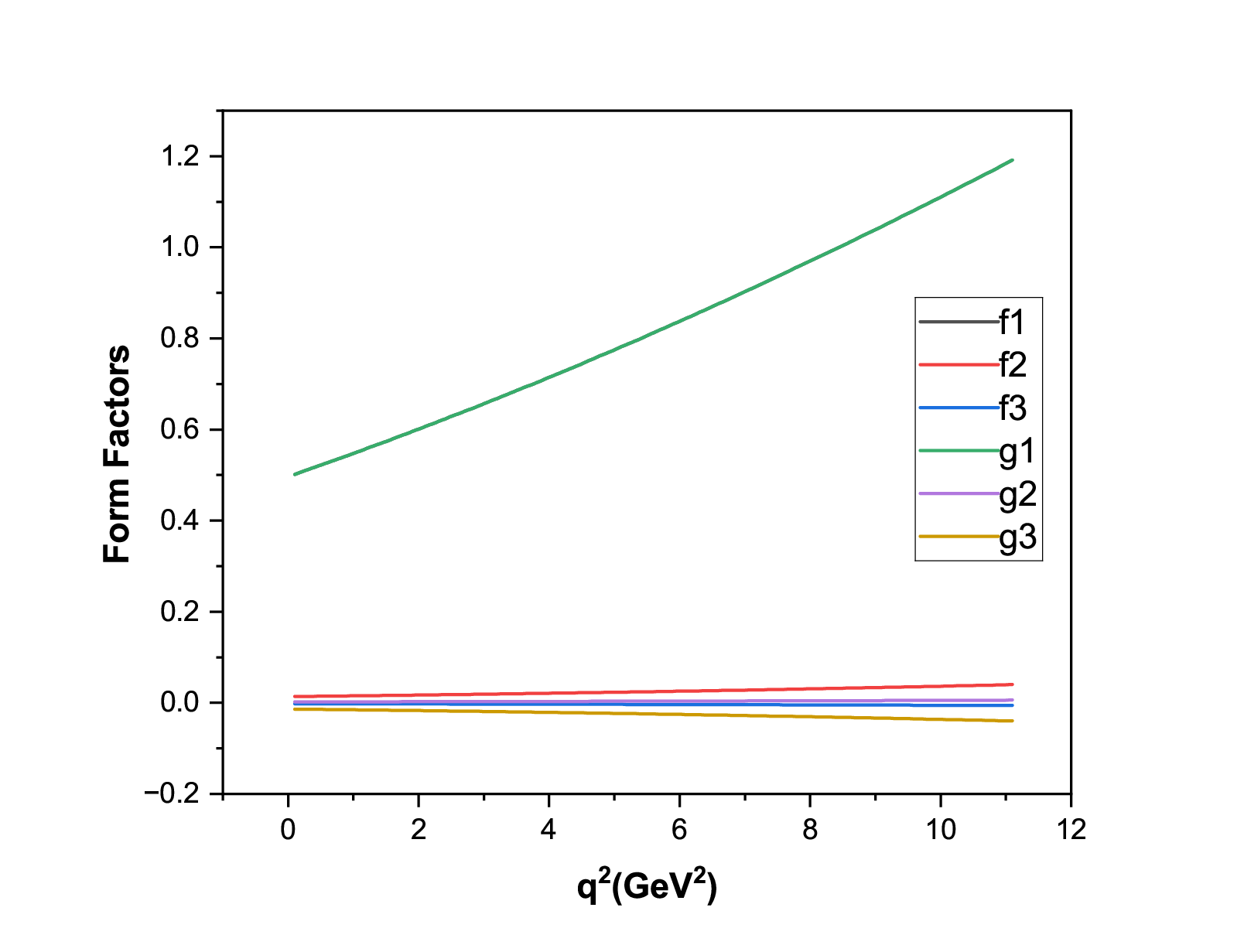}
\includegraphics[scale=0.3]{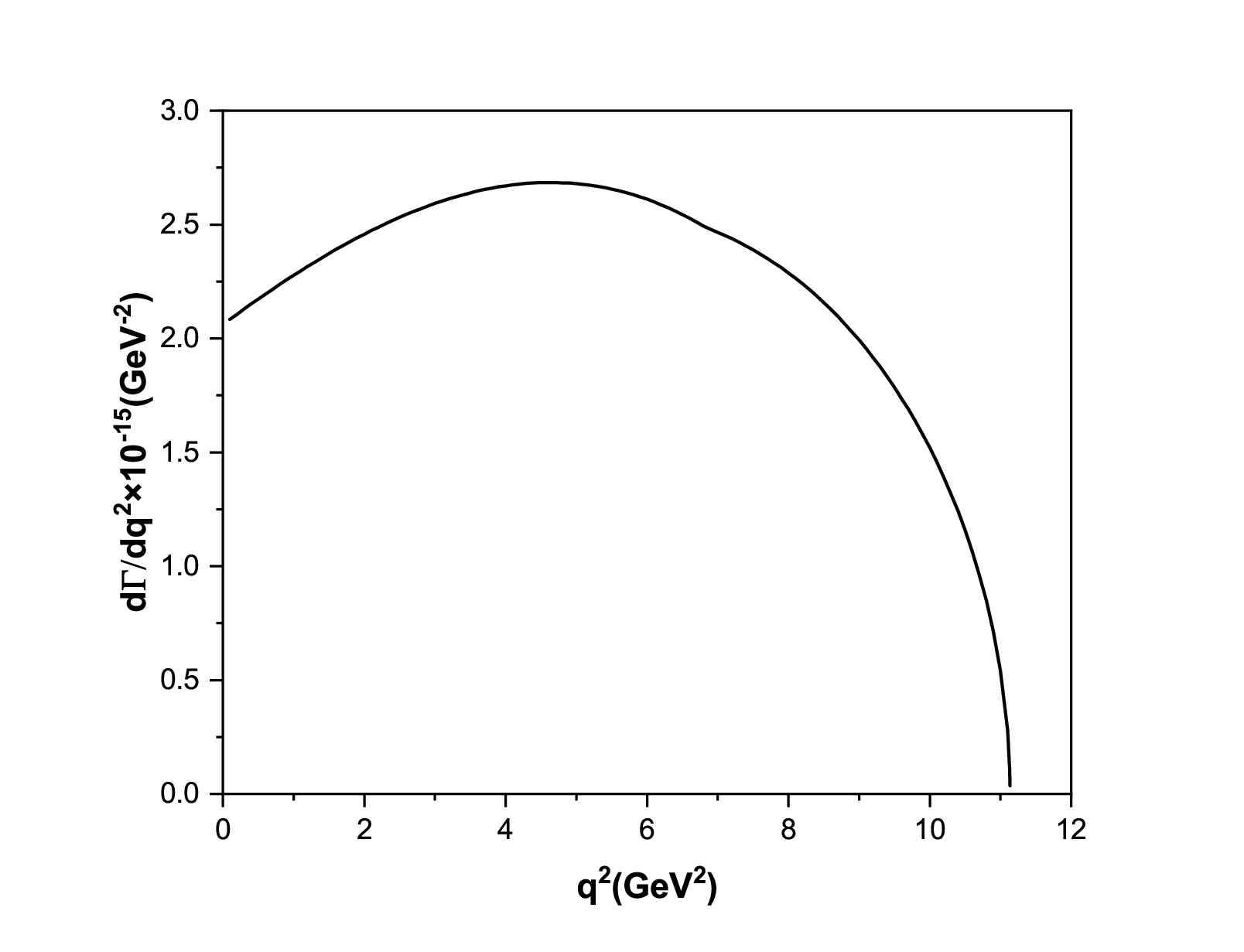}
\caption{\label{fig:1} $q^2$ dependency of six-form factors (left) and the variation of differential decay rate for $\Xi_b^0 \rightarrow \Xi_c^+e\bar{\nu}_e$ (right). }
\end{figure}

In this study, we calculated the ground-state masses and semileptonic decay properties of the $\Xi_{b}^{0}$ baryon using the HCQM. The model parameters and quark masses used in the numerical analysis are summarised in Table \ref{tab:table1}. The quark mass parameters are taken from previous studies \cite{Thakkar2020,Patel2025}, and we adjusted the model parameters $\beta$ and $V_0$ to reproduce the experimental ground state masses of the $\Xi_Q$ baryons. The predicted masses of the $\Xi_b$ and $\Xi_c$ baryons are $\Xi_b$ = 5.795 GeV and $\Xi_c$ = 2.468 GeV, respectively, which are in good agreement with the experimental results reported by the Particle Data Group \cite{pdg2024}. This agreement indicates that the chosen potential parameters and hyperfine corrections provide a reliable description of the internal dynamics of $\Xi_Q$ baryons, yielding physically meaningful baryon wavefunctions.
\begin{table}
\centering
    \caption{\label{tab:table1}Quark mass parameters and constants used in the calculations.}
    \begin{tabular}{cc}
    \noalign{\smallskip}\hline
    Parameter & Value\\
    \noalign{\smallskip}\hline
    ${m_{u}}$ & 0.33 GeV\\
    ${m_{s}}$ & 0.50 GeV\\
    ${m_{c}}$ & 1.55 GeV\\
    ${m_{b}}$ & 4.95 GeV\\
    $\beta$ & 0.2 $GeV^2$\\
    ${V_0}$ for $\Xi_b^0$ & -1.085 GeV\\
    ${V_0}$ for $\Xi_c^+$ &-1.020 GeV\\
    $x_0$ & 1.00 $GeV^{-1}$\\
    $\alpha_s(\mu_0 = 1 GeV)$ & 0.6\\
    $n_f$ & 4 \\
    \noalign{\smallskip}\hline
    \end{tabular}
\end{table}
Using the ground-state wavefunctions, we computed the Isgur-Wise function $\xi(\omega)$ for the $\Xi_b \rightarrow \Xi_c$ transition. The slope and convexity parameters obtained from the wavefunctions are $\rho^2 = 1.83$ and $c = 0.84$, respectively. Ref. \cite{Tazimi2021} reported the value of the slope $\rho^2 = 1.85$ and convexity parameters $c = 0.93$, which are in agreement with our obtained values. The Isgur-Wise function is then employed to evaluate all six transition form factors $F_{1,2,3}$ and $G_{1,2,3}$ that are relevant for semileptonic decay. As expected from the HQET, $F_1$ and $G_1$ which overlaps, dominate the kinematic region, whereas $F_{2,3}$ and $G_{2,3}$ are suppressed by $1/m_Q$ corrections. The small but nonvanishing values of $F_2$, $F_3$, $G_2$ and $G_3$ arise from $1/m_b$ and $1/m_c$ corrections, which are parameterised in our formalism through $\bar{\Lambda}/m_Q$. The $q^2$ behaviours of the six form factors in the allowed regions are illustrated in Fig. \ref{fig:1}. The calculated values of the form factors at $q^2=0$ and $q^2_{max}$ are listed in Table \ref{tab:table2}. It is observed that the form factors $f_1$ and $g_1$ exhibited nearly identical numerical values. Although the expressions for $f_1$ and $g_1$ are different, the invariant form factors entering these expressions satisfy $F_2 = G_2$ and $F_3 = -G_3$. $F_i$ and $G_i$ combine in such a way that $f_1$ and $g_1$ become nearly identical numerically. The obtained form factors are in agreement with form factors obtained using three different models in the Perturbative QCD and compared with other theoretical models in Ref. \cite{Rui2025}.\\
Using the obtained form factors and the helicity formalism, the total decay width for $\Xi_b^0 \rightarrow \Xi_c^+e\bar{\nu}_e$ is computed as $3.81 \times 10^{10} s^{-1}$. The decay width obtained for $\Xi_b^0 \rightarrow \Xi_c^+\tau\bar{\nu}_{\tau}$ transition is $1.24\times  10^{10} s^{-1}$, Table \ref{tab:table3} compares our results with existing theoretical predictions for decay width and branching ratio. The decay width reported by various theoretical predictions lies between $3.68 \times 10^{10} s^{-1}$ and $8.44\pm0.422 \times 10^{10} s^{-1}$. Our result lies toward the lower end of the predicted range and is consistent with several relativistic quark model calculations. Reducing the spread among theoretical predictions will require precise experimental measurements as well as refined theoretical calculations, including lattice QCD and higher-order HQET corrections. The estimated uncertainties were obtained by varying the quark mass parameters within a range of $\pm$0.05 GeV around their central values. This variation affects the baryon wavefunctions and masses obtained from the model, which in turn modifies the slope and curvature of the Isgur-Wise function. Consequently, the form factors and decay widths are also affected. The resulting decay rates are $3.81^{+1.36}_{-1.17} \times  10^{10} s^{-1}$ for $\Xi_b^0 \rightarrow \Xi_c^+e\bar{\nu}_e$ and $1.24^{+0.17}_{-0.17} \times  10^{10} s^{-1}$ for $\Xi_b^0 \rightarrow \Xi_c^+\tau\bar{\nu}_{\tau}$ transitions, respectively. The uncertainty in the semitauonic channel appears nearly symmetric due to its reduced phase space and larger lepton mass, whereas the semielectronic channel exhibits a more pronounced asymmetric behaviour. The larger and asymmetric uncertainty in the semielectronic channel arises from its sensitivity over the full kinematically allowed range of $q^2$, due to the negligible electron mass.

The branching ratio can be further obtained by $\mathcal{B} = \tau_{\Xi_b}\times\Gamma$ with $\tau_{\Xi_b} = 1.48 \times 10^{-12} s$ denoting the lifetime of $\Xi_b^0$ baryon \cite{pdg2024}. The predicted branching ratio obtained in the present work lies within the range of existing theoretical results, as shown in Table 3. The obtained branching ratios for $\Xi_b^0 \rightarrow \Xi_c^+e\bar{\nu}_e$ transition is $5.60^{+1.99}_{-1.72}\%$ and for $\Xi_b^0 \rightarrow \Xi_c^+\tau\bar{\nu}_{\tau}$ transition is $1.83^{+0.25}_{-0.25}\%$. Our result is closer to the lower end of the predicted range and is consistent with relativistic quark model calculations \cite{Faustov2018} and light front quark model \cite{Ke2024}. This agreement indicates that the present approach provides a reliable description of the semileptonic decay process. The variation among different theoretical predictions reflect differences in the treatment of non-perturbative QCD effects in various models. Therefore, precise experimental measurements of such decay modes are important to further constrain theoretical approaches and improve our understanding of heavy baryon dynamics. In addition, the Lepton Flavour universality (LFU) ratio is obtained as $\mathcal{R}_{\Xi_c} = \frac{\mathcal{B}_{\tau}}{\mathcal{B}_{e}} = 0.325$, which is in agreement with the value reported in Ref. \cite{Faustov2018}. As the experimental measurement of $\mathcal{R}_{\Xi_c}$ is not yet available, our predictions can provide additional hints for exploring LFU in the $b \rightarrow c\ell\bar{\nu}_{\ell}$ transition and may provide new insights into the $\mathcal{R}^{(*)}$ puzzle. \\
\begin{table}[h]
\centering
   \caption{\label{tab:table2} Form factors for $\Xi_b^0 \rightarrow \Xi_c^+e\bar{\nu}_e$ decay}
    \begin{tabular}{ccc}
    \noalign{\smallskip}\hline
    form factors & at $q^2=0$ & at $q^2_{max}$ \\
    \noalign{\smallskip}\hline
    $f_1$ 	&	0.496	&	1.193	\\
    $f_2$ 	&	0.013	&	0.040	\\
    $f_3$ 	&	-0.002	&	-0.005	\\
    $g_1$ 	&	0.496	&	1.193	\\
    $g_2$ 	&	0.002	&	0.005	\\
    $g_3$ 	&	-0.013	&	-0.040	\\
    \noalign{\smallskip}\hline
    \end{tabular}
\end{table}
Overall, the present analysis provides a consistent description of the semileptonic decay $\Xi_b^0 \rightarrow \Xi_c^+\,\ell\,\bar{\nu_\ell}$ within the HCQM framework combined with HQET. The dominance of the leading form factors, smooth kinematic behaviour, and agreement with existing theoretical predictions indicate that this approach offers reliable predictions for heavy baryon semileptonic transitions. The results presented herein may serve as useful theoretical benchmarks for future experimental measurements and refined theoretical studies.


\begin{table}[t]
\centering
    \caption{\label{tab:table3} Semileptonic decay width of $\Xi_b^0 \rightarrow \Xi_c^+\ell\bar{\nu}_\ell$ (in $10^{10} s^{-1}$) and branching ratio (in \%)}
    \begin{tabular}{ccc}
    \noalign{\smallskip}\hline
    Decay Width & Branching ratio & Ref.\\
    \noalign{\smallskip}\hline
    3.81	&	5.60	&	This work	\\
    3.68	&	5.41	&	\cite{Ke2024}	\\
    3.91	&	5.75	&	\cite{Faustov2018}	\\
    4.64	&	6.82	&	\cite{Ebert2006}	\\
    4.98	&	7.32	&	\cite{Albertus2005}	\\
    5.27	&	7.75	&	\cite{Ivanov1997}	\\
    5.30 	&	7.79	&	\cite{Cheng1996}	\\
    5.76	&	8.47	&	\cite{Hassanabadi2016} 	\\
    5.77	&	8.48	&	\cite{Zhao2020}	\\
    5.98	&	8.79	&	\cite{Ivanov2000} 	\\
    6.03	&	8.86	&	\cite{Zhao2018}	\\
    6.40    &   9.41    &   \cite{Rusetsky1997} \\
    6.83	&	10.04	&	\cite{Ivanov1999}	\\
    7.19	&	10.57	&	\cite{Singleton1991}	\\
    $7.46\pm0.38$	&	$10.97\pm0.56$	&	 \cite{Farhadi2023}	\\
    8.22	&	12.08	&	\cite{Hassanabadi2014}	\\
    $8.44\pm0.422$	&	$12.41\pm0.62$	&	\cite{Tazimi2021}	\\
    \noalign{\smallskip}\hline
    \end{tabular}
\end{table}

\section*{Acknowledgments}
 The author (K. Patel) acknowledges the financial assistance provided by the Anusandhan National Research Foundation (ANRF), Government of India, under the International Travel Support (ITS) scheme (File No. ITS/2025/004984), which enabled participation in the BARYONS 2025 conference.


\begin{thebibliography}{99}
\bibitem{pdg2024}S. Navas et al. (Particle Data Group), Phys. Rev. D 110, 030001 (2024). 			
\bibitem{Knapp1976}B. Knapp et al., Phys. Rev. Lett. 37, 882 (1976).
\bibitem{Bari1991a}G. Bari, M. Basile, G. Bruni et al., Nuovo Cim. A 104, 1787 (1991).
\bibitem{Bari1991b}G. Bari, M. Basile, G. Bruni et al., Nuovo Cim. A 104, 571 (1991).
\bibitem{Ammar2001} R. Ammar et al., Phys. Rev. Lett. 86, 1167 (2001).
\bibitem{Abazov2007}V. M. Abazov et al. [D0], Phys. Rev. Lett. 99, 052001 (2007).
\bibitem{Aaij1}R. Aaij et al. [LHCb], Phys. Rev. Lett. 128, no.16, 162001 (2022).
\bibitem{Aaij2}R. Aaij et al. [LHCb], Phys. Rev. D 103, no.1, 012004 (2021).
\bibitem{Sirunya2021} A. M. Sirunyan et al. [CMS], Phys. Rev. Lett. 126, no.25, 252003 (2021).
\bibitem{RAaij2023epjc} R. Aaij et al. (LHCb Collaboration), Eur. Phys. J. C 84, 3, 237 (2024).
\bibitem{Cheng1996} H. Y. Cheng and B. Tseng, Phys. Rev. D 53, 1457 (1996) [erratum: Phys. Rev. D 55, 1697 (1997)].
\bibitem{Hassanabadi2015}H. Hassanabadi, S. Rahmani, Few Body Syst. 56 10, 691 (2015).
\bibitem{Albertus2005}C. Albertus, E. Hern\'andez and J. Nieves, Phys. Rev. D 71, 014012 (2005) 	
\bibitem{Patel2025a} K. Patel, K. Thakkar, Eur. Phys. J. Plus 140, 452 (2025).
\bibitem{Bowler1998}K. C. Bowler et al., Phys. Rev. D 57, 6948 (1998).
\bibitem{Ivanov2000} M. A. Ivanov, J. G. Korner, V. E. Lyubovitskij, M. A. Pisarev and A. G. Rusetsky, Phys. Rev. D 61, 114010 (2000).
\bibitem{Ebert2006}D. Ebert, R. N. Faustov, V. O. Galkin, Phys. Rev. D 73, 094002 (2006).
\bibitem{Migura2006} S. Migura, D. Merten, B. Metsch and H. R. Petry, Eur. Phys. J. A 28, 55 (2006).
\bibitem{Ivanov1997}M. A. Ivanov, V. E. Lyubovitskij, J. G. Korner and P. Kroll, Phys. Rev. D 56, 348-364 (1997).
\bibitem{Faustov2018} R. N. Faustov and V. O. Galkin, Phys. Rev. D 98, no.9, 093006 (2018).
\bibitem{Zhao2020}Z. X. Zhao, R. H. Li, Y. L. Shen, Y. J. Shi and Y. S. Yang, Eur. Phys. J. C 80, no.12, 1181 (2020).
\bibitem{Neishabouri2025}Z. Neishabouri, K. Azizi, Phys. Rev. D 112 5, 054009 (2025).
\bibitem{Ivanov1999}M. A. Ivanov, J. G. K\"{o}rner, V. E. Lyubovitskij,  A. G. Rusetsky, Phys. Rev. D 59, 074016 (1999).
\bibitem{Ke2024} H. W. Ke, G. Y. Fang and Y. L. Shi, Phys. Rev. D 109, no.7, 073006 (2024).
\bibitem{Zhao2018}Z. X. Zhao, Chin. Phys. C 42, no.9, 093101 (2018).
\bibitem{Cardarelli1999} F. Cardarelli and S. Simula, Phys. Rev. D 60, 074018 (1999).
\bibitem{Thakkar2020}K. Thakkar, Eur. Phys. J. C 80 (10), 926 (2020).
\bibitem{Ferraris1995} M. Ferraris, M. M. Giannini, M. Pizzo, E. Santopinto, L. Tiator, Phys. Lett. B 364, 231 (1995).			
\bibitem{Giannini1983}M. M. Giannini,  Nuovo Cim. A 76, 455 (1983).			
\bibitem{Garcilazo2007}  H. Garcilazo, J. Vijande and A. Valcarce, et al., J. Phys. G 34, 961 (2007).
\bibitem{Patel2025}K. Patel, K. Thakkar, Int. J. Theor. Phys. 64, 5, 129 (2025).			
\bibitem{majethiya2008a}A. Majethiya, B. Patel and P. C. Vinodkumar, Eur. Phys. J. A 38, 307 (2008).

\bibitem{Gutsche2015}T. Gutsche, M. A. Ivanov, J. G. K\"{o}rner, V. E. Lyubovitskij, P. Santorelli and N. Habyl, Phys. Rev. D 91, 074001 (2015).
\bibitem{Patel2026} K. Patel, K. Thakkar, Eur. Phys. J. A 62, 65 (2026).
\bibitem{Falk1993} A. Falk, M. Neubert, Phys. Rev. D 47, 2982 (1993).
\bibitem{Isgur1991}N. Isgur and M. B. Wise, Nucl. Phys. B 348, 276 (1991).
\bibitem{Georgi1990} H. Georgi, B. Grinstein, and M. B. Wise, Phys. Lett. B 252, 456 (1990).
\bibitem{Neubert1994}M. Neubert, Phys. Rep. 245, 259 (1994).
\bibitem{Bialas1993}P. Bialas, J. G. K\"{o}rner, M. K\"{a}rmer, K. Zalewski, Z. Phys. C 57, 115 (1993).
\bibitem{Faustov2016} R. N. Faustov and V. O. Galkin, Phys. Rev. D 94, 7, 073008 (2016).


\bibitem{Rui2025} Z. Rui, Z.-T. Zou, Y. Li, and Y. Li, Phys. Rev. D 111, 113006 (2025).
\bibitem{Hassanabadi2014}	H. Hassanabadi, S. Rahmani, S. Zarrinkamar, Phys. Rev. D 90, 074024 (2014).
\bibitem{Rusetsky1997}	A. G. Rusetsky, M. A. Ivanov, J. G. K\"{o}rner, V. E. Lyubovitskij, arXiv:hep-ph/9710524 [hep-ph].
\bibitem{Singleton1991}R. Singleton, Phys. Rev. D 43, 2939 (1991).	
\bibitem{Farhadi2023}M. Farhadi et al. Eur. Phys. J. A  59, 171 (2023)

\bibitem{Tazimi2021}N. Tazimi, P. Sadeghi Alavijeh, Adv. High Energy Phys. 7713697 (2021).
\bibitem{Hassanabadi2016}H. Hassanabadi and S. Rahmani, Eur. Phys. J. Plus 2, 34 (2016)


\end{thebibliography}
\end{document}